\documentclass[twocolumn,english]{revtex4-1}
\usepackage[T1]{fontenc}
\usepackage[latin9]{inputenc}
\usepackage{amsmath}
\usepackage{amssymb}
\usepackage{graphicx}
\usepackage{babel}
\usepackage{xcolor}
\usepackage{stix}
\usepackage[pdftex,colorlinks]{hyperref}
\definecolor{darkblue}{cmyk}{1,0,0,0.8}
\definecolor{darkred}{cmyk}{0,1,0,0.7}
\hypersetup{anchorcolor=black,
  citecolor=darkblue, filecolor=darkblue,
  menucolor=darkblue,pagecolor=darkblue,urlcolor=darkblue,linkcolor=darkblue}
\begin{document}

\title{Temporal dissipative solitons in time-delay feedback systems}

\author{Serhiy Yanchuk$^{1}$, Stefan Ruschel$^{1}$, Jan Sieber$^{2}$,
Matthias Wolfrum$^{3}$}

\address{$^{1}$Institute of Mathematics, Technical University of Berlin,
Strasse des 17 Juni 136, 10623 Berlin, Germany}

\address{$^{2}$Harrison Building, North Park Road, CEMPS University of Exeter,
Exeter, EX4\textrm{ }4QF, UK}

\address{$^{3}$Weierstrass Institute, Mohrenstrasse 39, 10117 Berlin, Germany}
\begin{abstract}
Localized states are a universal phenomenon observed in spatially
distributed dissipative nonlinear systems. Known as dissipative solitons,
auto-solitons, spot or pulse solutions, these states play an important
role in data transmission using optical pulses, neural signal propagation,
and other processes. While this phenomenon was thoroughly studied
in spatially extended systems, temporally localized states are gaining
attention only recently, driven primarily by applications from fiber
or semiconductor lasers. Here we present a theory for temporal dissipative
solitons (TDS) in systems with time-delayed feedback. In particular,
we derive a system with an advanced argument, which determines the
profile of the TDS. We also provide a complete classification of the
spectrum of TDS into interface and pseudo-continuous spectrum. We
illustrate our theory with two examples: a generic delayed phase oscillator,
which is a reduced model for an injected laser with feedback, and
the FitzHugh-Nagumo neuron with delayed feedback. Finally, we discuss
possible destabilization mechanisms of TDS and show an example where
the TDS delocalizes and its pseudo-continuous spectrum develops a
modulational instability.
\end{abstract}
\maketitle
Solitons have been known as a physical phenomenon from the early 19th
century \cite{ScottRussell1844}. They are commonly associated with
spatially localized states in conservative spatially extended systems,
such as the Korteweg-de Vries or the nonlinear Schrödinger equation
and possess remarkable properties such as preservation of localization
and shape after collisions. Beyond the ``classical'' conservative
solitons, localized states were also observed in earlier works on
non-conservative chemical and physiological systems, see \cite{Purwins2010}
and references therein. 

Interest in localized solutions of non-conservative and non-integrable
systems has grown rapidly since the early 1990s \cite{Kerner1994,Rosanov2002,Akhmediev2005,Akhmediev2008,Ackemann2009,Purwins2010,Grelu2012,Parra-Rivas2013}.
These states have been called dissipative solitons (DS). In contrast
to conservative solitons, DS are stable objects (attractors), which
emerge due to a nonlinear balance between energy gain and loss \cite{Grelu2012}.
DS have been discovered in spatially extended systems modeled by partial
differential equations in optics \cite{Kerner1994,Barland2002,Akhmediev2005,Ackemann2009,Jang2013,Grelu2012,Bednyakova2015},
biological systems \cite{Kerner1994,Heimburg2005,Lautrup2011,VillagranVargas2011},
plasma physics \cite{Kerner1994,Tur1992} and other fields \cite{Rotermund1991}. 

Recent experimental and theoretical results report that DS are also
possible in systems with time-delayed feedback that do not include
explicit spatial variables \cite{Vladimirov2005,Marconi2014,Marino2014,Marconi2015,Garbin2015a,Romeira2016,Semenov2018,Brunner2018,Liu2018,Schelte2018}.
In these systems the time delay is larger than the other timescales
and the DS are temporally localized. Their natural relation to spatially
localized states can be seen in a spatio-temporal representation of
the dynamics of time-delayed systems as done in \cite{Giacomelli1996,Yanchuk2017}.
In this representation the pulse is localized within the \emph{delay
line}. For example, in a ring laser this delay line corresponds physically
to the ring cavity, where the optical pulse is localized \cite{Vladimirov2005}. 

Examples of systems exhibiting temporal DS (TDS) 
include opto-electronic 
setups such as mode-locked lasers
with saturable absorber \cite{Vladimirov2005,Marconi2014,Schelte2018}, coupled
broad-area semiconductor resonators \cite{Genevet2008}, vertical-cavity
surface-emitting lasers with delays \cite{Marconi2015}, as well as neuronal
models \cite{Romeira2016} or bistable systems with feedback 
\cite{Marino2014,Semenov2018}. 
Although localized states have been reported
mainly in one dimension, two-dimensional TDS have been found as well
for a system with two feedback loops \cite{Brunner2018}. In this
case the lengths of the delays were significantly different. Then
one can associate one spatial dimension to each delay line, thus representing
the temporal dynamics using a two-dimensional spatial representation
\cite{Yanchuk2014,Yanchuk2015a}. Localized states can have different
forms. For instance, they can be composed of several pulses, known
as soliton molecules or bound states \cite{Marconi2015,Liu2018,Puzyrev2017}.
Experimental and theoretical methods to control the nucleation or
cancellation of TDS have been introduced in \cite{Garbin2015a,Romeira2016}. 

\begin{figure}
\includegraphics[width=1\columnwidth]{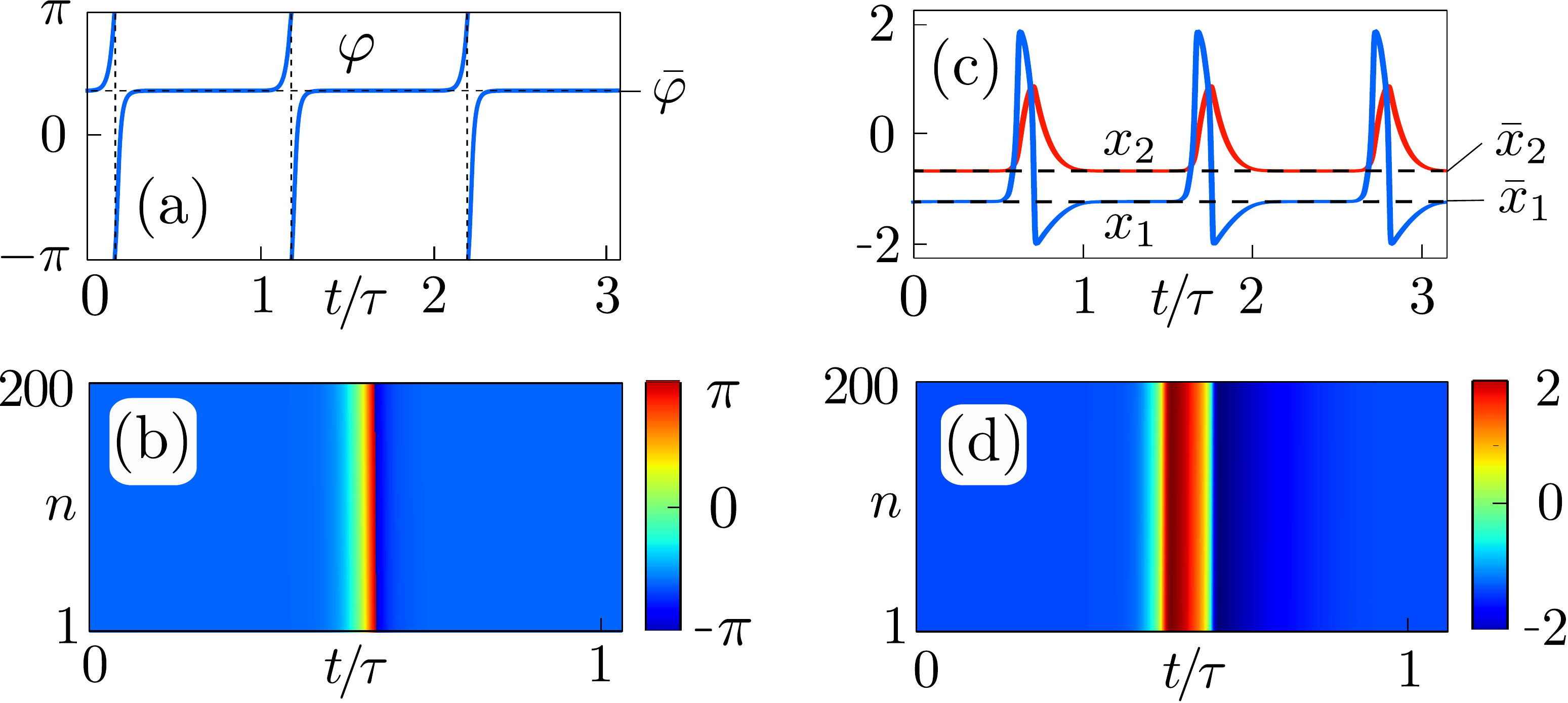}\caption{\label{fig:Examples-of-localized} Examples of temporal dissipative
solitons (TDS) in the delayed phase oscillator (\ref{eq:phaseosc})
(a,b) and FHN system (\ref{eq:FHN}) (c,d). Panels (a,c) show the
time profiles $\varphi(t)$ and $x(t)$ and (b,d) their spatio-temporal
representations. The spatio-temporal representation shows the solutions
$\varphi(t)$ in (b) (and $x_{1}(t)$ in (d)) as color plot with respect
to the pseudo-spatial variable (delay-line) along the horizontal axis
($t/\tau$ mod $T/\tau$) and the pseudo-temporal variable (number
of round-trips) along the vertical axis $(n=\left[t/T\right])$ \cite{Giacomelli1996,Yanchuk2017}.
Parameter values: (a,c) $d=0.9,\kappa=1,\tau=40,$ (b,d) $a=0.7,b=0.8,\kappa=0.1,\varepsilon=0.08,\tau=100$. }
\end{figure}

Considering the importance of TDS in systems with delayed feedback,
their variety and broadness of applications, there is a need for a
unifying theory describing basic properties of TDS. In this Letter,
we outline such a theory for TDS with a stable equilibrium \emph{background
state }(see Fig.~\ref{fig:Examples-of-localized} for typical time
profiles) for general systems with delayed feedback of the form
\begin{equation}
\dot{x}(t)=f(x(t),x(t-\tau)),\label{eq:1}
\end{equation}
where $x(t)\in\mathbb{R}^{n}$ is a variable describing the state
of the system, $\tau$ is the large feedback delay, and $f(\cdot,\cdot)$
is a nonlinear function determining the dynamics. 

We present two ingredients that enable TDS to emerge in systems (\ref{eq:1}),
and introduce an equation describing the TDS time profile. Using the
largeness of time delay $\tau$, we describe the spectrum of Floquet
multipliers of TDS. This spectrum consists of two parts. The first
is the \emph{pseudo-continuous spectrum} (PCS), determined entirely
by (but not equal to) the spectrum of its background state. We provide
an explicit expression for the PCS when the time-delayed feedback
has rank 1 and a simple description for PCS computation otherwise.
The second part is a point (or \emph{interface}) spectrum, for which
we provide an asymptotic approximation that is independent of the
large delay $\tau$ and hence can be evaluated numerically (see \cite{Sieber2014}). 
The obtained
results predict possible destabilization mechanisms of TDS. We specify
these mechanisms and conclude by showing an example of delocalization
of TDS and the development of a modulational instability. 

Examples of TDS are shown in Fig.~\ref{fig:Examples-of-localized}
for the delayed phase oscillator 
\begin{equation}
\dot{\varphi}=d-\sin\varphi+\kappa\sin\left(\varphi(t-\tau)-\varphi\right),\label{eq:phaseosc}
\end{equation}
and the FitzHugh-Nagumo (FHN) neuron with delayed feedback 
\begin{align}
\dot{x}_{1} & =x_{1}-(x_{1}^{3}/3)-x_{2}+\kappa x_{1}(t-\tau),\nonumber \\
\dot{x}_{2} & =\varepsilon(x_{1}+a-bx_{2}).\label{eq:FHN}
\end{align}
	System (\ref{eq:phaseosc}) is  a reduced model for a general injected Ginzburg-Landau equation with delayed feedback \cite{Garbin2015a} (see
Fig.~\ref{fig:Examples-of-localized} for parameters).

We observe that TDS are periodic solutions with a period $T$ slightly
larger than the time delay $\tau$. We denote $T=\tau+\delta$ where
$\delta\ll\tau$ will remain bounded as $\tau$ gets large. As Fig.~\ref{fig:Examples-of-localized}
shows, the solutions spend most of the time close to a constant stationary
state $\bar{x}$, which we call the \emph{background}. 

\emph{Conditions for the emergence of TDS and profile equation. }The
first ingredient is the existence of a background equilibrium $\bar{x}$
that is stable for arbitrary long delay $\tau$. The equilibrium $\bar{x}$
satisfies $f(\bar{x},\bar{x})=0$. It is stable if all roots $\lambda$
of the characteristic equation $\det(\lambda\text{I}-A_{0}-B_{0}\exp(-\lambda\tau))=0$
have negative real parts \cite{Hale1977}. Here $A_{0}=\partial_{1}f(\bar{x},\bar{x})$
and $B_{0}=\partial_{2}f(\bar{x},\bar{x})$ are Jacobians of the function
$f$ with respect to the first and second argument, respectively,
evaluated at $\bar{x}$. Interestingly, stability of the background
for long delays implies its stability for arbitrary positive delays
$\tau$ including small and zero delay \footnote{If $\bar{x}$ is unstable for some delay $\tau_{0}>0$, then there
must be a bifurcation for the larger value $\tau_{b}>\tau_{0}$ where
$\lambda=i\omega$ is purely imaginary, and, hence, $i\omega\text{I}-A-Be^{i\omega\tau_{b}}=0$.
The latter equality implies that $i\omega\text{I}-A-Be^{i\omega\tau_{k}}=0$
for all delays $\tau_{k}=\tau_{b}+2\pi k/\omega$ with arbitrary integer
$k$, thus, contradicting to the stability for long delays.}. Explicit stability criteria for large delays $\tau$ are given in
\cite{Lichtner2011}. 

The second ingredient refers to the time profile $s(t)$ of the TDS.
Using its $T=\tau+\delta$-periodicity, we find that $s(t)$ satisfies
(\ref{eq:1}) if and only if
\begin{equation}
\dot{s}(t)=f(s(t),s(t+\delta))\label{eq:delta}
\end{equation}
since $s(t-\tau)=s(t-\tau+T)=s(t+\delta)$. In the resulting \emph{profile
equation} (\ref{eq:delta}), where the large time delay is replaced
by a finite positive time shift $\delta$, the TDS appears as a family
of periodic solutions with long periods that for some positive $\delta=\delta_{h}$
approaches a \emph{connecting orbit} (also called homoclinic solution)
$s_{h}(t)$ to $\bar{x}$. We recall that a connecting orbit satisfies
$s_{h}(t)\to\bar{x}$ for $t\to\pm\infty$, i.e. it approaches the
background $\bar{x}$ forward and backward in time. Clearly, such
an orbit cannot exist for negative $\delta$ because the background
$\bar{x}$ is stable in (\ref{eq:1}). Another reason for the positive
sign of $\delta_{h}$ is the causality principle \cite{Yanchuk2017}
which implies that the period of a stable TDS is larger than the time-delay
$\tau$.

The homoclinic solution $s_{h}(t)$ of the profile equation (\ref{eq:delta})
with $\delta=\delta_{h}$ implies the appearance of TDS in system
(\ref{eq:1}) for large delays $\tau$ in the following way. Considering
$\delta$ as a parameter in (\ref{eq:delta}), the general theory
for connecting orbits \cite{Shilnikov2001,Homburg2010} guarantees that for $\delta$
close to $\delta_{h}$, the profile equation possesses a family of
periodic solutions $s_{\delta}(t)$ with periods $T_{\delta}$ approaching
infinity as $\delta\to\delta_{h}$. These periodic solutions converge
to the connecting orbit with infinite period as $\delta\to\delta_{h}$.
Using the periodicity, we have $s_{\delta}(t+\delta)=s_{\delta}(t+\delta-T_{\delta})=s_{\delta}(t-\tau)$
with $\tau=T_{\delta}-\delta$. Hence, $s_{\delta}(t)$ solves (\ref{eq:1})
with $\tau=T_{\delta}-\delta$. Since $T_{\delta}$ goes to infinity,
the branch of periodic solutions $s_{\delta}(t)$ of the original
system (\ref{eq:1}) also exists for the large time delay $\tau=T_{\delta}-\delta$
with $\tau\to\infty$, $\delta\to\delta_{h}$. Moreover, the solutions
$s_{\delta}$ are close to the connecting orbit, and hence, they are
TDS.

In short, the main ingredients leading to TDS are: \\
(A) A background equilibrium $\bar{x}$ that is stable for large and,
hence, also for arbitrary positive delays. \\
(B) The profile equation (\ref{eq:delta}) possesses a connecting
orbit to $\bar{x}$ for some positive value $\delta_{h}$. The period
of the TDS is then approximately $T\approx\tau+\delta_{h}$ for large
delays. 

The profile equation (\ref{eq:delta}) is a differential equation
with an advanced argument. This is in contrast to the profile equations
for spatial DS \cite{Kerner1994,Rosanov2002,Akhmediev2005,Akhmediev2008,Ackemann2009,Purwins2010,Grelu2012},
which are ordinary differential equations.

\begin{figure}[t]
\includegraphics[width=1\columnwidth]{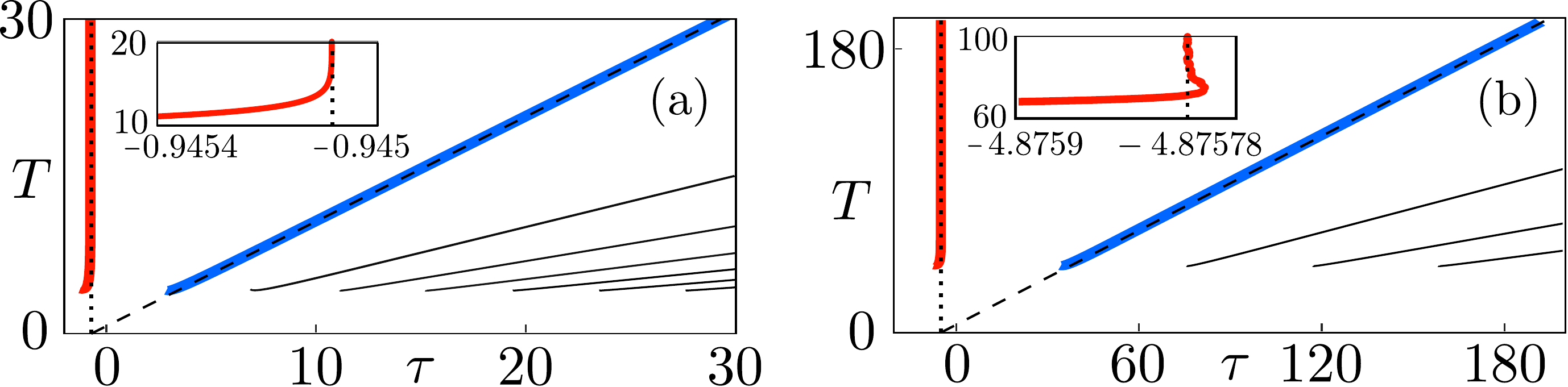}

\caption{\label{fig:Bifurcation-diagrams} Branches of periodic solutions:
(a) delayed phase oscillator (\ref{eq:phaseosc}); (b) FHN system
(\ref{eq:FHN}); period $T$ versus delay $\tau$. The primary branch
of TDS (solid blue curves) has the asymptotic period $T=\tau+\delta_{h}$
(dashed line). The branch reappears for negative delays $-\delta=\tau-T$
(red lines) and limits to the connecting orbit of the profile equation
(\ref{eq:delta}) with $\delta\to\delta_{h}$ and $T\to\infty$ (dotted
line). Higher harmonic TDS branches (black lines) correspond to the
branches reappearing with time-delays $\tau+kT(\tau)$ (multiple solitons
per delay interval). Other parameters: (a) $d=0.9,\kappa=0.9$ (b,d)
$a=0.7,b=0.8,\kappa=0.1,\varepsilon=0.08$. }
\end{figure}

The bifurcation diagram in Fig.~\ref{fig:Bifurcation-diagrams} illustrates
the relation between the solutions of the profile equation (red branch)
and the TDS solutions (blue branch), showing the periods as a function
of the time-delay $\tau$. One can clearly see the asymptotic behavior
$T\approx\tau+\delta_{h}$ for the period along the blue primary stable
branch of TDS. The branches are related by the general reappearance
rule $\tau_{k}=\tau+kT(\tau)$, see \cite{Yanchuk2009}, where $k=0$
corresponds to the blue branch, $k=-1$ to the red, and $k>2,3,\dots$
to the higher harmonic branches (black). The defining feature for
TDS is that the period along the red branch diverges, and that the
periodic solutions approach the connecting orbit $s_{h}(t)$ as $\tau\to-\delta_{h}$.

\begin{figure}
\includegraphics[width=1\columnwidth]{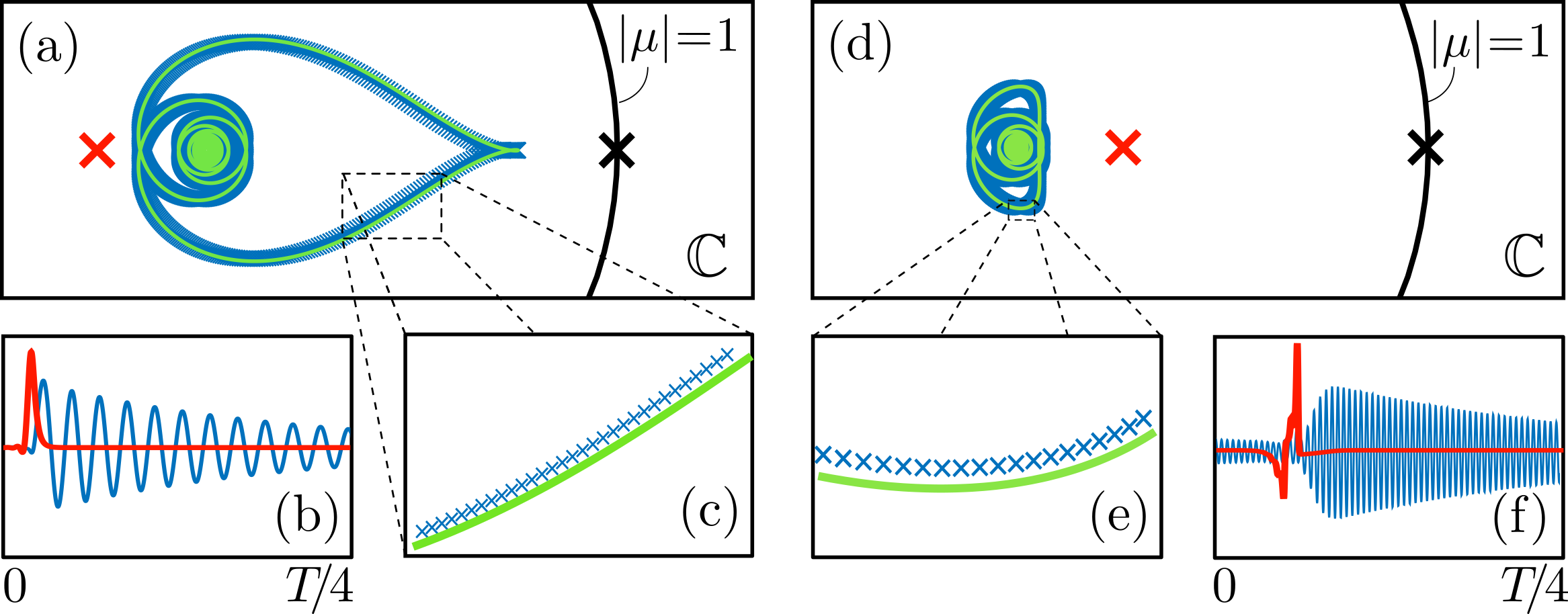}

\caption{Spectrum and eigenfunctions of TDS: (a)-(c) delayed phase oscillator
(\ref{eq:phaseosc}); (d)-(f) FHN system (\ref{eq:FHN}). Panels (a),(d)
with zoomed parts in panels (c), (e) show numerically computed multipliers
(crosses) and the approximating curves (\ref{eq:PCS-phase}) and (\ref{eq:PCS-FHN})
for the PCS (green curves). Interface spectrum (red and black crosses)
can be computed using the Evans function (\ref{eq:is}). Eigenfunctions
in panels (b), (f): Localized profiles (red) correspond to interface
spectrum; non-localized profiles (blue) correspond to PCS. Parameters
for (a)-(c): $d=0.9,\kappa=0.9,\tau=200$; for (d)-(f): $a=0.7,b=0.8,\kappa=0.1,\varepsilon=0.08,\tau=1000$.
\label{fig:spectrum}}
\end{figure}

\emph{Spectrum of TDS and mechanisms for its destabilization.} Next
we describe the spectrum of TDS, which determines the stability, possible
bifurcations and destabilization scenarios of TDS. We show that the
spectrum has two parts: \emph{pseudo-continuous} (PCS) and \emph{interface}
spectrum, see Fig.~\ref{fig:spectrum}. The PCS is determined by
the background while the interface spectrum consists of usually only
few relevant multipliers that are determined by the profile properties. 

To determine the spectrum, system (\ref{eq:1}) is linearized around
the TDS solution $s_{\delta}(t)$: 
\begin{equation}
\dot{y}(t)=A(t)y(t)+B(t)y(t-\tau),\label{eq:lin}
\end{equation}
where $A(t)=\partial_{1}f(s_{\delta}(t),s_{\delta}(t+\delta))$ and
$B(t)=\partial_{2}f(s_{\delta}(t),s_{\delta}(t+\delta))$. Taking
into account the properties of TDS, the coefficients $A(t)$ and $B(t)$
are most of the time exponentially close to $A_{0}$ and $B_{0}$,
respectively, except for intervals of length of order 1 where the
TDS is different from the background.

The linearized system (\ref{eq:lin}) determines the dynamics of small perturbations 
$y(t)=x(t)-s_{\delta}(t)$ around the TDS. Its coefficients $A(t)$ and $B(t)$ are $T$-periodic, 
therefore, accordingly to the Floquet theory \cite{Hale1977}, 
special solutions $y(t)$ of this system with the property $y(t+T)=\mu y(t)$ are 
eigenfunctions while the corresponding complex numbers $\mu$ are multipliers. 
In particular, the multipliers are related to the Lyapunov exponents $\lambda$ as
$\mu=\exp{(\lambda T)}$.
For stable TDS all multipliers have $|\mu|<1$, except the trivial one $\mu=1$ corresponding
to the time-shift. 
 
When searching for the multipliers and eigenfunctions, using the 
equality $y(t-\tau)=\mu^{-1}y(t-\tau+T)=\mu^{-1}y(t+\delta)$, we obtain from (\ref{eq:lin})
the following eigenvalue problem
\begin{equation}
\dot{y}(t)=A(t)y(t)+\mu^{-1}B(t)y(t+\delta),\ \ y(t+T)=\mu y(t).\label{eq:EP}
\end{equation}
Our next goal is to find approximations of the solutions $y(t)$ and $\mu$
of (\ref{eq:EP}) for large $T$. In the following, we present the results leaving the
technical details in the Supplemental Material (SM). 
The following characteristic equation
\begin{equation}
\det\Delta(\mu,\rho)=\det\left(\rho I-A_{0}-\mu^{-1}e^{\rho\delta}B_{0}\right)=0, \label{eq:rho}
\end{equation}
which determines the stability of the profile equation (\ref{eq:EP}) at the background,  plays an
important role. 

One distinguishes two types of multipliers $\mu$: interface spectrum,
for which the characteristic equation (\ref{eq:rho}) possesses no
purely imaginary roots $\rho=i\omega$, and PCS, where (\ref{eq:rho})
has such purely imaginary roots. 

\emph{Interface spectrum.}
The multipliers from the interface spectrum are given as roots of the following equation:
$
\det E(\mu)=0,\label{eq:is}
$
where  $E$ is a $k_s\times k_s$ matrix, and $k_s$ is the number of stable roots $\rho_j$ (with negative real parts) 
of (\ref{eq:rho}). All elements of the matrix $E$ are defined independently of the large
delay $\tau$ or period $T$. Its explicit form is given in SM, and it 
has the same structure as the Evans functions
for localized solutions in spatially extended systems \cite{Sandstede2002,Akhmediev2005}.
An algorithm for computing the interface spectrum using the presented
theory and DDE-Biftool  is available as a demo in \cite{Sieber2014}.
Figure~\ref{fig:spectrum} shows examples of the interface spectrum
(red and black crosses in panels (a) and (d)). According to the construction
in the SM the corresponding eigenfunctions $y(t)$ in Fig.~\ref{fig:spectrum}
are localized at the interface and decay exponentially to zero in
the background region of the TDS (red profiles in panels (b) and (f)). 

\emph{Pseudo-continuous spectrum }(\emph{PCS})\emph{ }(blue crosses
in Fig.~\ref{fig:spectrum})\emph{ }is given by multipliers $\mu$,
for which the characteristic equation (\ref{eq:rho}) has purely imaginary
roots $\rho_{c}=i\omega$. Substituting $\rho=i\omega$ in (\ref{eq:rho}),
we obtain $\det\Delta(\mu,i\omega)=\det\left(i\omega I-A_{0}-\mu^{-1}e^{i\omega\delta}B_{0}\right)=0$.
This relation determines a curve $\mu(\omega)$ in the complex plane
(green curves in Fig.~\ref{fig:spectrum}), along which the multipliers
$\mu$ of the PCS accumulate. For scalar systems this curve has the
form $\mu(\omega)=e^{i\omega\delta}B_{0}/(i\omega-A_{0})$, which
gives for (\ref{eq:phaseosc})
\begin{equation}
\mu(\omega)=\kappa e^{i\omega\delta_{h}}/(i\omega+cos\bar{\varphi}+\kappa).\label{eq:PCS-phase}
\end{equation}
 In systems with more variables, the equation $\det\Delta(\mu,i\omega)=0$
is a polynomial of degree $\mathrm{rank}\,B_{0}$ in $\mu^{-1}$.
In the FitzHugh-Nagumo system (\ref{eq:FHN}) the feedback is scalar
($\mathrm{rank}\,B_{0}=1$), giving 
\begin{equation}
\mu(\omega)=\kappa(\varepsilon b+i\omega)e^{i\omega\delta_{h}}/(\varepsilon+(\bar{x}_{1}^{2}+i\omega-1)(\varepsilon b+i\omega)).\label{eq:PCS-FHN}
\end{equation}

The imaginary root $\rho_{c}=i\omega$ of Eq.~(\ref{eq:rho}) implies
that the eigenfunction $y(t)$ of the corresponding multiplier $\mu(\omega)$
is a multiple of $e^{i\omega t}$ far from the interface soliton
and hence, in contrast to the eigenfunctions of the interface spectrum,
it is not localized (blue profiles in Figs.~\ref{fig:spectrum}(b),(f)). 

The presented theory allows a detailed study of TDS in any system
with delayed feedback of the form (\ref{eq:1}). While delay systems
with large delay are typically characterized by high dimensional dynamics,
our approach of separating the large timescale of delay 
from the short timescale of the soliton interface allows to find the
soliton profile and the interface spectrum from the desingularized
equations
independently
of the large delay. Indeed, the interface spectrum describes the linear
response with respect to variations of the shape and position of the
soliton interface. Corresponding instabilities are induced by isolated
multipliers and can be studied within the classical framework of low-dimensional 
systems, leading to e.g. period-doubled or quasiperiodically
modulated TDS. Moreover, on the level of the profile
equation (\ref{eq:delta}), the bifurcations of the TDS can be related
to the theory of homoclinic bifurcations \cite{Shilnikov2001,Homburg2010}. Note
that classical codimension-two homoclinic bifurcations (e.g. orbit
flip, inclination flip, or Shilnikov type) appear here
already under the variation of a single control parameter of (\ref{eq:1}),
since the time shift $\delta$ appears as an additional unfolding
parameter in (\ref{eq:delta}).
However, as soon as the
background equilibrium ceases to be hyperbolic the high dimensional
nature of the system comes into play. Similarly to the critical continuous
spectrum at background instabilities of spatially extended systems,
PCS approaching the unit circle describes the corresponding phenomenon
for TDS. 

\begin{figure}
\includegraphics[width=1\columnwidth]{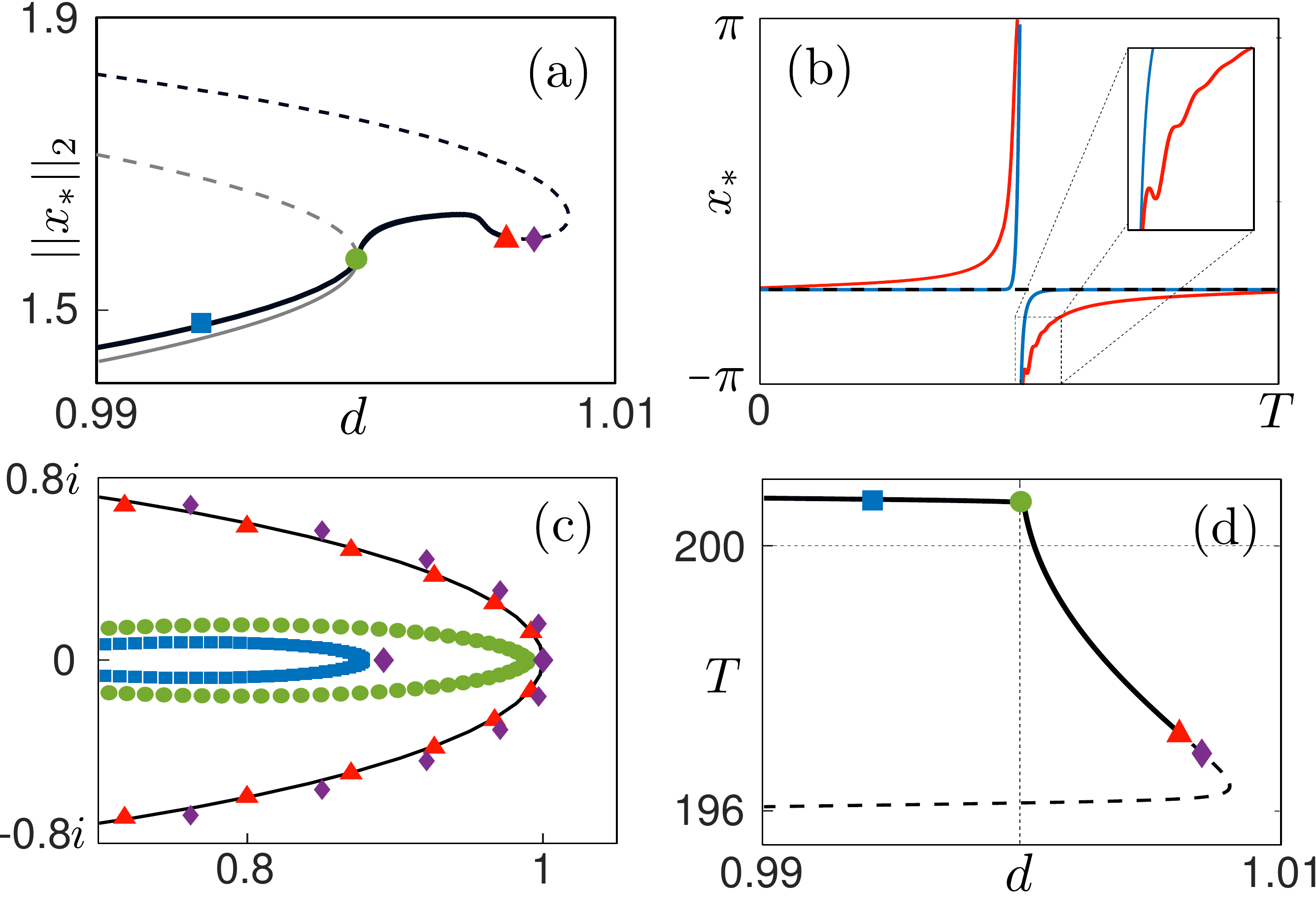}

\caption{\label{fig:example} Delocalization and development of modulational
instability of a TDS in system (\ref{eq:phaseosc}). (a) solution
branches of the background steady state (gray) and the periodic solution
(black) versus the excitability parameter $d$. Numerically obtained
Floquet spectra (c) and profiles (b) of selected periodic solutions,
indicated by points of corresponding color in (a). Panel (d) shows
period versus $d$. Other parameters $\kappa=0.9$, $\tau=200$.}
\end{figure}

We conclude with an example showing that in such situations specific
new dynamical scenarios have to be expected. In Fig.~\ref{fig:example}
we study numerically the destabilization of TDS in the phase oscillator
system (\ref{eq:phaseosc}) as the excitability parameter $d$ changes.
With increasing $d$, the background equilibrium $\bar{\varphi}$,
given by $d=\sin\bar{\varphi}$, disappears in a saddle-node bifurcation
at $d=1$, see gray solid and dashed lines in Fig.~\ref{fig:example}(a) for the 
stable and unstable branches, respectively.

Despite the disappearance of the background, there is still a stable
localized periodic solution, spending most of its period in the region
where the background equilibrium has vanished. 
Such a state exists within a small parameter interval 
of length of order $1/\tau$, see black solid line between the green point and the red triangle in Fig.~\ref{fig:example}(a). 
Strictly speaking, it is no more a TDS, as the ''ghost'' of the saddle-node
equilibrium serves as the new background for this state. Indeed, after
the background equilibrium vanishes, orbits still slow down in the
region of the phase space of the profile equation where the equilibrium
formerly existed. If the time spent in the ghost region is longer
than the time-delay, the ghost region can effectively serve as the
background. 

Let us discuss what happens with the localized solution along this branch. First of all, 
at $d=1$ the PCS of the TDS touches the imaginary axis (green points in Fig.~\ref{fig:example}(c)) and the localization of the phase soliton
becomes no longer exponential. Following this periodic branch further, 
the period becomes smaller than the delay (see Fig.~\ref{fig:example}(d)) 
and the solution loses its stability. 
This instability involves a large number of multipliers, which originate
from the former PCS and create a destabilization scenario similar
to a modulational instability. The change of the spectrum is illustrated in 
Fig.~\ref{fig:example}(c)  with increasing parameter $d$. In particular, one can see that many multipliers around the trivial one become unstable shortly after crossing the threshold.
Interestingly, the type of the destabilizations of the TDS and the background are different: modulational for the TDS while uniform for the background.
 Finally, the soliton branch turns back into
the region $d<1$, now as a highly unstable soliton solution, which
is attached to an unstable background equilibrium.

For this and other TDS destabilization scenarios our theory provides
a systematic framework, which can be considered as a substantial extension
of the classical theory for dissipative solitons in spatially extended
systems. Similar to modulational instability, other types of destabilizations could be predicted 
and studied such as e.g. oscillatory when the PCS destabilizes at nonzero frequencies, or uniform. 
The theory can be also used for studying the effect of noise on TDS, since it provides a tool for 
quantifying the projection of the noise on the most sensitive modes. 
The proposed theory can be extended to other localization 
phenomena in systems with delayed feedback such as e.g. localized fronts \cite{Nizette2004}.

\begin{acknowledgments}
Financial support acknowledgments. SY: Deutsche Forschungsgemeinschaft (DFG, project 411803875); JS: EPSRC via grants EP/N023544/1 and EP/N014391/1, and European Union's Horizon 2020 research and innovation programme under the Marie Sklodowska-Curie grant agreement No 643073; 
SR: IRTG 1740/TRP 2015/50122-0, funded by DFG/FAPESP; MW and SR: Deutsche Forschungsgemeinschaft (DFG, project number 163436311, SFB 910).
\end{acknowledgments}

\bibliographystyle{apsrev4-1}
%

\end{document}